\RequirePackage{fix-cm}
\let\oldvec\vec
\documentclass[smallextended]{svjour3}       
\usepackage{graphicx}
\let\vec\oldvec
\smartqed  
\usepackage{natbib}
\usepackage{amsfonts}
\usepackage{float}
\usepackage{alltt,calc,bm}
\usepackage{mathtools}
\usepackage{dsfont,color,graphicx,mathrsfs,enumerate,setspace,lipsum}
\usepackage{rotating}
\usepackage{xcolor}
\usepackage{ulem}
\usepackage{lineno}

 \journalname{}
\begin{document}

\title{Dating and localizing an invasion from post-introduction data and a coupled reaction-diffusion-absorption model
\thanks{This research was funded by a PhD grant INRA-R{\'e}gion PACA (Emplois Jeunes Doctorants 2016-2019), the HORIZON 2020 XF-ACTORS Project SFS-09-2016 and the INRA-DGAL project 21000679. We thank DGAL, Anses, SRAL, FREDON, LNR-LSV and certified laboratories for data collection and data availability. We thank Afidol for their endorsement in the PhD grant.}
}
%


\titlerunning{Dating and localizing an invasion}        

\author{C. Abboud \and
        O. Bonnefon \and
        E. Parent \and 
        S. Soubeyrand
}


\institute{C. Abboud \at BioSP, INRA, 84914 Avignon, France\\
              \email{candy.abboud@inra.fr}
           \and
          O. Bonnefon \at BioSP, INRA, 84914 Avignon, France
           \and
           E. Parent \at AgroParisTech, UMR 518 Math. Info. Appli., Paris, France
           \at    INRA, UMR 518 Math. Info. Appli., Paris, France
           \and 
           S. Soubeyrand \at BioSP, INRA, 84914 Avignon, France
}

\date{Received: date / Accepted: date}

\maketitle
\begin{abstract}
Invasion of new territories by alien organisms is of primary concern for environmental and health agencies and has been a core topic in mathematical modeling, in particular in the intents of reconstructing the past dynamics of the alien organisms and predicting their future spatial extents. Partial differential equations offer a rich and flexible modeling framework that has been applied to a large number of invasions. In this article, we are specifically interested in dating and localizing the introduction that led to an invasion using mathematical modeling, post-introduction data and an adequate statistical inference procedure. We adopt a mechanistic-statistical approach grounded on a coupled reaction-diffusion-absorption model representing the dynamics of an organism in an heterogeneous domain with respect to growth. Initial conditions (including the date and site of the introduction) and model parameters related to diffusion, reproduction and mortality are jointly estimated in the Bayesian framework by using an adaptive importance sampling algorithm. This framework is applied to the invasion of \textit{Xylella fastidiosa}, a phytopathogenic bacterium detected in South Corsica in 2015, France.


\end{abstract}
 
\section{Introduction}
\label{intro}
Biological invasions have long been an important topic for biologists and mathematicians because of their impact on the environment, indigenous species, and health of humans, animals and plants \citep{andow1990spread,andow1993spread,baker1991continuing,hengeveld_dynamics_1989,kermack_contributions_1991,richardson1991determinants,simberloff1989insect,anderson_transmission_1996,shigesada_biological_1997,weinberger1978asymptotic}. Biological invasions are generally viewed as the result of a process with four stages: arrival, establishment, spread and concentration \citep{reise2006aliens,vermeij1996agenda}. 
Each stage of the invasion process has been a core topic in mathematical modeling since the mid-twentieth century \citep{fisher_wave_1937,mollison_spatial_1977,okubo_diffusion_1980,shigesada_modeling_1995,skellam_random_1951}, and better understanding processes governing invasions is chiefly relevant for improving surveillance and control strategies. In particular, extensive researches have been conducted in the intents of reconstructing the past dynamics \citep{boys2008bayesian,roques2016using,soubeyrand_parameter_2014} of alien species and predicting their future spatial extents \citep{chapman2015inventory,peterson2003temporal}. In this context, partial differential equations offer a rich and flexible framework that has been applied to a large number of invasions \citep{gatenby_reaction-diffusion_1996,lewis1993,murray2002mathematical,okubo2002diffusion,turchin1998quantitative}. Even though a partial differential equation does not describe all the processes involved in an ecological dynamics, it can help in understanding its important properties and inferring its major components, such as dates and sites of invasive-species introductions.

Consider as an example the emergence of \textit{Xylella fastidiosa} (Xf), a phytopathogenic bacterium detected in South Corsica, France, in 2015 and currently present in a large part of this island \citep{denance2017several,soubeyrand2018inferring}. This plant pathogen has the potential to cause a major sanitary crisis in France, typically like in Italy, where a large number of infected olive trees dried and died,  causing serious damages to olive cultivation. To avoid such a situation, the French General Directorate of Food (DGAL) implemented enhanced control and surveillance measures after the first {\it in situ} detection of Xf in Corsica, which generated a data set consisting of a spatio-temporal point pattern (i.e. the locations and dates of plant samples) marked by a binary variable indicating the result of the diagnostic test (i.e. indicating if the plant sample is positive or negative to Xf).

In this example, only post-introduction data are available (i.e. data collected over a temporal window covering a period after the introduction time), and we precisely propose in this article an approach for estimating the date and the site of the introduction using such observational partial data. It has however to be noted that estimating the introduction point from post-introduction data requires the estimation of the propagation characteristics of the invasive species (and {\it vice versa}) because these characteristics link the introduction and the observations. Thus, in this paper, we aim at jointly estimating the date and site of the introduction, and other parameters related to growth, dispersal and death that govern the post-introduction dynamics. 

Such a joint estimation was proposed by \citet{soubeyrand_parameter_2014} with a simple reaction-diffusion model and was applied to simulated data. It was developed in a mechanistic-statistical framework that has often  been used to describe and infer ecological processes. This framework combines a mechanistic model for the dynamics of interest, a probabilistic model for the observation process and a statistical procedure for estimating model parameters \citep{berliner2003physical,lanzarone2017bayesian,roques2011,soubeyrand_2009_a,soubeyrand_2009b,wikle_2003a,wikle_2003b}. 
We adapted this framework for dating and localizing the introduction of an invasive species by taking into account spatial heterogeneities in growth and mortality. Precisely, we built a mechanistic model yielding the probability for the invasive species to occupy any spatial units at any time. This spatio-temporal function, with values in $[0,1]$, satisfies (i) a reaction-diffusion equation that describes the spread of the alien species in a sub-domain of the study domain and (ii) a diffusion-absorption equation that describes the dispersal and the death of the alien species in the complementary sub-domain. Typically, the partition into the two sub-domains can be determined by environmental variables affecting the growth and mortality of the invasive species (e.g. host/non-host environment, low/high winter temperature, and presence/absence of nutrients). In addition, our model assumes that there is only one introduction point (in time and space) that governs the emergence of the invasive species and that eventual other introduction points have negligible effects on the dynamics.
Estimation of model parameters, including the time and the location of the introduction, is carried out in a Bayesian framework with the adaptive multiple importance sampling algorithm \citep[AMIS;][]{bugallo2015adaptive,cornuet2012adaptive}. Contrary to the MCMC algorithm that is often used in the mechanistic-statistical framework (see references above), AMIS can be easily parallelized and its tuning parameters are automatically adapted across the algorithm iterations. Finally, the two sub-domains (where the reaction-diffusion and diffusion-absorption equations are defined) are obtained by thresholding a spatial variable. The threshold value is determined with a selection criterion. Four criteria are considered: the Bayesian information criterion \citep[BIC;][]{schwarz1978estimating}, two versions of the deviance information criteria \citep[DIC;][]{gelman2004bayesian,spiegelhalter2002bayesian} and a predictive information criterion \citep[IC;][]{ando2011predictive}.

In the Xf case study, the two sub-domains are defined by thresholding the average of the minimum daily temperature in January and February, the two coldest months of the year in Corsica. Indeed, winter temperature has been inferred as an important environmental factor governing the dynamics of Xf and the level of disease severity caused by Xf \citep{costello_environmental_2017,feil_effects_2003,feil_temperature-dependent_2001,henneberger2003effects,purcell1977cold,purcell1980environmental}. For instance, isolines for the average minimum daily temperature in January have been shown to be quite consistent with regions in the United States that are exposed to different levels of severity of the Pierce's disease of grape caused by Xf \citep{anas2008effect}.

The paper is structured as follows. The hierarchical modeling framework coupling a partial differential equation and a Bernoulli observation is described in Section \ref{sec:processmodel}. Bayesian inference for parameter estimation relying on an adaptive MCMC algorithm and four criteria for model selection are also presented in this methodological section. The results obtained from surveillance data for Xf in the case study (Corsica) are detailed in Section \ref{sec:application}. In Section \ref{sec:discussion}, we summarize and discuss our work.

\section{The mechanistic-statistical approach}
\label{sec:method}

\subsection{Process model}
\label{sec:processmodel}

Models based on parabolic partial differential equations have often been used to describe biological invasions \citep{skellam_random_1951,shigesada_modeling_1995,shigesada_biological_1997,okubo_diffusion_1980}. Here, we are interested in the invasion of a pathogen, that spreads in a domain $\Omega$ included in $\mathbb{R}^2$. We assume that there is only one single introduction point in time and space that triggered the invasion and that eventual subsequent introductions have negligible effects on the dynamics and are therefore not incorporated into the model. Furthermore, to account for spatial heterogeneity in the reproduction regime of the pathogen, we divide $\Omega$ into two sub-domains, say $\Omega_{1}$ and $\Omega_{2}$, such that $\Omega=\Omega_{1}\cup\Omega_{2}$, $\Omega_{1}\cap\Omega_{2}=\emptyset$ and different growth terms apply to $\Omega_{1}$ and $\Omega_{2}$.

More formally, the spread of the pathogen is described by a coupled model governing the probability $u(t,\mathbf{x})$ of a host located at site $\mathbf{x}=(x_1,x_2)\in\Omega$ to be infected at time $t$. This model is grounded on two particular types of parabolic partial differential equations: (i) a reaction-diffusion equation in $\Omega_1$ where the growth is logistic \citep{verhulst1838notice} and (ii) a diffusion-absorption equation in $\Omega_2$ where only dispersal and death events occur.
The probability $u(t,\mathbf{x})$ satisfies:
\begin{equation}\label{xf}
\begin{cases}
\dfrac{\partial u}{\partial t}=D \Delta u + b u\left(1-\dfrac{u}{K}\right) \mathds{1}(\mathbf{x}\in\Omega_{1}) -\alpha u \mathds{1}(\mathbf{x}\in\Omega_{2}), & \,\,t\geq\tau_0,\,\,\mathbf{x}\in\Omega,\\
\nabla u(t,\mathbf{x}).n(\mathbf{x})=0, & \,\,t\geq\tau_0,\,\,\mathbf{x}\in\partial\Omega,\\
u(\tau_0,\mathbf{x})=u_0(\mathbf{x})\geq 0, & \,\,\mathbf{x}\in\Omega,\\
\end{cases}
\end{equation} 
where $D>0$ is the diffusion coefficient; $b$ corresponds to the intrinsic growth rate of the pathogen; $K\in(0,1]$ is a plateau for the probability of infection (i.e. an analog to the carrying capacity of the environment); $\Delta=\dfrac{\partial^2}{\partial x_1^2}+\dfrac{\partial^2}{\partial x_2^2}$ is the 2-dimensional diffusion operator of Laplace; $\mathbf{x}\mapsto\mathds{1}(\mathbf{x}\in\Omega_{i})$ is the characteristic function taking the value $1$ if $\mathbf{x}\in\Omega_{i}$ and $0$ otherwise; $\tau_0\in\mathbb{R}$ is the introduction time of the pathogen. As explained in the introduction, the sub-domains $\Omega_1$ and $\Omega_2$ are defined by thresholding a spatial function, say $T$, with the threshold value $\tilde T$ that is hold fixed: $\Omega_1=\Omega_1(T,\tilde T)=\{{\mathbf x}\in\Omega:T({\mathbf x})> \tilde T\}$ and $\Omega_2=\Omega_2(T,\tilde T)=\{{\mathbf x}\in\Omega:T({\mathbf x})\le \tilde T\}$.

In our framework, the initial condition $u_0$ models the introduction of the pathogen in the study domain. Here, the introduction represents the initial phase of the outbreak corresponding to the arrival of the pathogen and its local establishment. Thus, $u_0$ is not expressed as a Dirac delta function but as a kernel function centered around the central point of the introduction $\mathbf{\tilde{x}}_0=(\tilde{x}_0,\tilde y_0)\in\Omega$. More precisely, the probability of a host at $\bf x$ to be infected at $\tau_0$ satisfies:
\begin{equation}\label{initialcondition}
\centering
u_0({\bf x})=p_0 \exp \left( -\dfrac{\lVert \mathbf{x}-\mathbf{\tilde{x}}_0 \rVert^2}{2\sigma^2}\right)
\end{equation}
where $p_0$ is the infection probability at $(\tau_0,\mathbf{\tilde{x}}_0)$, $\sigma^2=\frac{r_0^2}{q}$, $q$ is the 0.95-quantile of the $\chi^2$ distribution with two degrees of freedom, and $r_0$ is the {\it radius} of the kernel. Thus, at $\tau_0$, if we neglect border effects, 95\% of the infected plants are located within the ball with center $\mathbf{\tilde{x}}_0$ and radius $\mathit{r}_0$. Assuming in addition reflecting conditions on the boundary $\partial\Omega$ of $\Omega$, the system of equations \eqref{xf} is well-posed \citep{evans_partial_1998}. In addition, by constraining $p_0$ in $[0,K]$, the principle of parabolic comparison \citep{protter1967maximum} implies that the solution of \eqref{xf} is also in the interval $[0,K]$.

\subsection{Data model}
\label{sec:datamodel}

Let $t_i\in{\mathbb R}$ denote the sampling time of host $i\in\{1,\ldots,I\}$, $I\in{\mathbb N}^*$, $\mathbf{x}_i\in\Omega$ its location and $Y_i\in\{0,1\}$ its sanitary status observed at time $t_i$ (1 for infected, 0 for healthy). Conditionally on $u$, $T$ and $\{(t_i,\mathbf{x}_i):1\le i \le I\}$, the sanitary statuses $Y_i$, $i\in\{1,\ldots,I\}$, are assumed to be independent random variables following Bernoulli distributions with success probability $u(t_i,\mathbf{x}_i)$: 

\begin{equation}\label{datamodel}
\centering
Y_i \mid u,T,\{(t_i,\mathbf{x}_i):1\le i \le I\} \underset{\text{indep.}}{\sim}\text{Bernoulli}(u(t_i,\mathbf{x}_i)),
\end{equation}
where $u$ depends on parameters $D$, $b$, $K$, 
$\alpha$, $\tau_0$, $\mathbf{\tilde{x}}_0$, $\mathit{r}_0$, $p_0$ and $\tilde{T}$. 

{\it Remark.}
This simple data model could be modified to account for factors classically encountered in epidemiology, e.g. false-positive and false-negative observations, and spatial and temporal dependencies not accounted for in the process model.

\subsection{Parameter estimation with an adaptive importance sampling algorithm}
\label{sec:estimation}

Inference about the parameter vector $\Theta=(D,b,K,\alpha,\tau_0,\mathbf{\tilde{x}}_0,\mathit{r}_0,p_0)$ is made in the Bayesian framework, which technically consists in assessing the posterior distribution $[\Theta|Y]$ of $\Theta$ conditional on sanitary statuses $Y=\{Y_i:1\leq i\leq I\}$. The parameter $\tilde{T}$ will be treated later in section \ref{sec:selection} via model selection. Philosophically, a posterior probability is to be interpreted as a coherent judgment quantifying a subjective degree of uncertainty \citep{Lin2006}.

In what follows, we will keep using Gelfand's bracket notations for probability distributions \citep{gelfand1990sampling}. The posterior distribution of the unknown, hereafter dubbed $\Theta$, is derived by Bayes' rule:

\begin{equation*}
\centering
[\Theta|Y] = \frac{[Y|\Theta]\times[\Theta]}{[Y]},
\end{equation*}
where 

$[Y|\Theta]$ is the conditional distribution of the data $Y$ given the unknown $\Theta$ (i.e. the likelihood function of the model) that satisfies (using Equation \eqref{datamodel}):
\begin{equation}
\label{likelihood}
[Y|\Theta] 
 = \prod_{i=1}^I u(t_i,\mathbf{x}_i)^{Y_i}(1-u(t_i,\mathbf{x}_i))^{1-Y_i};
\end{equation} 
$[\Theta]$ is the prior distribution of $\Theta$ that depends on the application and that will be specified in Section \ref{sec:application}; the distribution of $Y$, $[Y]=\int[Y|\Theta][\Theta]d\Theta$, may be a formidable integral, depending of the dimension of the unknown $\Theta$. However, modern Bayesian algorithms \citep{Brooks03} avoid its computation by making recourse to Monte Carlo techniques only based on the unnormalized probability function $[Y|\Theta]\times[\Theta]$. Yet, the computation of $[Y|\Theta]$ itself requires the value of the solution $u$ of Equation \eqref{xf} for any valid parameter vector $\Theta$. This equation admits a unique solution for any fixed and valid $\Theta$, but cannot be solved analytically. Hence, we make recourse to a standard finite-element method with the software  \verb"Freefem++"  \citep{hecht2012new}; see Section \ref{sec:solution}.

For the mechanistic-statistical model defined above, the posterior distribution $[\Theta|Y]$ cannot be expressed analytically due to its intractable normalizing constant, but one can draw a sample under this distribution using an adequate algorithm for Bayesian inference. The so-called posterior sample $[\Theta|Y]$ is then used to numerically characterize all that we know about $\Theta$ after data assimilation. Here, we use the adaptive multiple importance sampling \citep[AMIS;][]{cornuet2012adaptive} algorithm, that consists of iteratively generating parameter vectors under an adaptive proposal distribution and assigning weights to the parameter vectors. To design efficient importance sampling algorithms, the auxiliary proposal distribution should be chosen as close as possible to the posterior distribution. However, the posterior distribution being unknown, the crucial choice of the proposal is a difficult task \citep{gelman1996efficient,roberts1997weak}. The main aim of the AMIS algorithm is to overcome this difficulty by tuning the coefficients of the proposal distribution picked in a parametric family of distributions, generally the Gaussian one, at the end of each iteration.

In this framework, at each iteration, new coefficient values for the proposal distribution are determined using the current weighted posterior sample \citep{bugallo2015adaptive}, then the posterior sample is augmented by generating new replicates from the newly tuned proposal distribution and the weights of the cumulated posterior sample are recomputed. The algorithm can be described as follows:

\begin{enumerate}
\item Set initial values $\mu_0$ and $\Sigma_0$ for the mean vector and the variance matrix of the multi-normal proposal distribution $\mathcal{N}(\mu_0,\Sigma_0)$, whose probability density function is denoted by $\Theta\rightarrow g_{\mu_0,\Sigma_0}(\Theta)$.
\item At iteration $m=1,\cdots,M$,
\begin{enumerate}
\item Generate a new sample $\{\Theta_m^l:l=1\cdots,L\}$ from the proposal distribution $\mathcal{N}(\mu_{m-1},\Sigma_{m-1})$.
\item Compute the un-normalized importance weights for the new sample as in Equation \eqref{eq:w1}, and update the un-normalized weights for the previously generated samples as in Equation \eqref{eq:w2}:
\begin{align}
\label{eq:w1}
 \tilde{w}_m^l & = \dfrac{[Y|\Theta_m^l]\times[\Theta_m^l]}{\dfrac{1}{m}\sum\limits_{j=1}^{m}g_{\mu_{j-1},\Sigma_{j-1}}(\Theta_m^l)}, \,\, l=1,\cdots,L\\
\label{eq:w2}
\tilde{w}_\varepsilon^l & = \dfrac{[Y|\Theta_\varepsilon^l]\times[\Theta_\varepsilon^l]}{\dfrac{1}{m}\sum\limits_{j=1}^{m}g_{\mu_{j-1},\Sigma_{j-1}}(\Theta_\varepsilon^l)}, \,\, \varepsilon=1,\cdots,m-1,\,\,l=1,\cdots,L.
\end{align}
\item Normalize the weights:
\begin{equation*}
w_\varepsilon^l=\dfrac{\tilde{w}_\varepsilon^l}{\sum\limits_{i=1}^{m}\sum\limits_{j=1}^{L}\tilde{w}_i^j},\,\,\varepsilon=1,\cdots,m,\,\,l=1,\cdots,L.
\end{equation*}
\item Adapt coefficient values for the next proposal distribution as follows:
\begin{align*}
\mu_m & =\sum\limits_{l=1}^{L}\sum\limits_{\varepsilon=1}^{m}w_\varepsilon^l \Theta_\varepsilon^l \\
\Sigma_m & =\sum\limits_{l=1}^{L}\sum\limits_{\varepsilon=1}^{m}w_\varepsilon^l ( \Theta_\varepsilon^l-\mu_\varepsilon)(\Theta_\varepsilon^l-\mu_\varepsilon)^t.\hspace*{10cm}
\end{align*}
\end{enumerate}
\end{enumerate}
The AMIS algorithm provides a weighted posterior sample $\{\{\Theta_m^l,w_m^l\}_{l=1}^L\}_{m=1}^M$ of size $ML$, which provides an empirical approximation of the posterior distribution $[\Theta|Y]$. Conditions leading to the convergence in probability of the posterior mean of any function (integrable with respect to the posterior distribution) of the parameters are described in \cite{cornuet2012adaptive} and are satisfied in our case.

\subsection{Choice of $\tilde T$ with a model selection procedure}
\label{sec:selection}

Implementation constraints concerning the partition of the study domain which depends on the threshold $\tilde T$, led us to proceed by two separate steps: (i) to infer model parameters for different fixed values of $\tilde T$ and, then, (ii) to select the value of $\tilde T$ having the largest support of data (this amounts to selecting a model within a class of models characterized by $\tilde T$).
Thus, for each element $\tilde T_a$ in $\{\tilde T_1,\ldots,\tilde T_A\}\subset\mathbb R^A$, $A\in\mathbb N^*$, we carried out the estimation procedure described in Section \ref{sec:estimation} by instancing $\tilde T$ at the value $\tilde T_a$ and letting it fixed. Then, the best value of $\tilde T$ is chosen by minimizing some criteria classically used for model selection: here we rely on the Bayesian Information criterion \citep[BIC;][]{schwarz1978estimating}, two Deviance information criteria \citep[DIC;][]{spiegelhalter2002bayesian,gelman2004bayesian} and a predictive Information Criterion \citep[IC;][]{ando2011predictive}. We use different selection criteria in order to report the variability of the selected $\tilde T$ when different hypotheses are made about which the best model is, if any.

The BIC satisfies:
\begin{equation}
\text{BIC} =-2\log[Y|\hat{\Theta}]+k \log I,
\end{equation}
where $I$ is the sample size, $k$ is the number of model parameters (in our setting, $k$ is the same for all the models), and $\hat{\Theta}$ is the maximum likelihood estimate of the parameter vector $\Theta$ in the support $\mathcal S(\Theta;\tilde T_a)$ of $\Theta$ defined by the prior distribution (in our setting, this support depends on the fixed value $\tilde T_a$ of $\tilde T$):
$$\hat{\Theta}=\underset{\Theta\in\mathcal S(\Theta;\tilde T_a)}{\text{argmax}}[Y|\Theta].$$ 
The DIC satisfies:
\begin{equation}
\text{DIC} =\mathcal{\bar D}+p_\text{eff},
\end{equation}
where $\bar{\mathcal{D}}$ is the posterior mean of the deviance $\mathcal{D}(\Theta)=-2\log[Y|\Theta]+C$ (where $C$ is a constant that cancels out when one compares different models) and $p_\text{eff}$ is the effective number of parameters of the model. The difference in the two versions of the DIC considered here lies in the calculation of $p_\text{eff}$. In the first version proposed by \citet{spiegelhalter2002bayesian},
\begin{equation}
p_\text{eff}=p_\mathcal{D}=\mathcal{\bar D}-\mathcal{D}(\bar{\Theta}), 
\end{equation}
where $\bar{\Theta}$ is the posterior mean of $\Theta$: $\bar{\Theta}=\mathds{E}[\Theta|Y]$. In the second version proposed by \citet{gelman2004bayesian},
\begin{equation}
p_\text{eff}=\frac{1}{2}\mathds{V}(\mathcal{D}(\Theta)|Y),
\end{equation}
where $\mathds{V}(\mathcal{D}(\Theta)|Y)$ is the posterior variance of $\mathcal{D}(\Theta)$.
The IC of \citet{ando2011predictive}, which is supposed to solve over-fitting issues, satisfies:
 \begin{equation}
\text{IC} =\bar{\mathcal{D}}+2p_\mathcal{D}:=3\bar{\mathcal{D}} -2\mathcal{D}(\bar \Theta).
\end{equation}
In practice, the different terms appearing in the four criteria, namely $\hat\Theta$, $\bar\Theta$, $\bar{\mathcal{D}}$ and $\mathds{V}(\mathcal{D}(\Theta)|Y)$, are replaced by their empirical values using the weighted posterior sample $\{\{\Theta_m^l,w_m^l\}_{l=1}^L\}_{m=1}^M$ provided by the application of the AMIS algorithm.

\subsection{Numerical equation solving}
\label{sec:solution}

For the application, computations were carried out with the software  \verb"Freefem++"  ~\citep{hecht2012new}. A Finite Element Method was used. The nonlinearity has been treated with a Newton-Raphson algorithm applied to the variational formulation of Equation~(\ref{xf}), by instancing the criterion of convergence at the value $10^{-10}$. The solution was approximated by a piecewise linear and continuous function. The time resolution was based on an adaptive step size using a backward Euler method. Supplementary Figure S1 shows the spatial discretization composed of 4791 nodes that has been used in the application in Section \ref{sec:application}. With this mesh, the average computation time for one simulation is $55$ seconds. 
We explored the effect of the spatial discretization by comparing the numerical solutions of the equation obtaind with the 4791-nodes mesh and with a finer mesh composed of 10703 nodes. The solutions were computed for the set of parameters corresponding to the posterior maximum (Supplementary material S4 shows the time continuous dynamics for this set of parameters). Supplementary Figure S2 shows very close simulation results for both meshes. Moreover, we investigated the numerical error of system \ref{xf} by using the indicator, norm $||u||_{H^2}$ which is classically considered to  control the $H^1$-error~\citep{ALL2008}. Using the mesh composed of 4791 nodes leads to a numerical error around $0.02$ corresponding to a satisfying accuracy for our application. 
 
\section{Application to the dynamics of \textit{Xylella fastidiosa} in South Corsica}
\label{sec:application}

\subsection{Surveillance data}
\label{sec:data}

For this application, we use spatio-temporal binary data on the presence of {\it Xylella fastidiosa} (Xf) collected in South Corsica, France, from July 2015 to May 2017. Over this period, approximately 8000 plants were sampled, among which 800 have been diagnosed as infected \citep[with a real-time PCR technique;][]{denance2017several}. Available data for all the sampled plants are their spatial coordinates, their sampling dates and their health statuses. Coordinates and health statuses at the sampling times are shown in Figure \ref{fig:1}. 

\subsection{Model specifications}
\label{sec:modelspec}

\begin{figure}[t]
\centering
\includegraphics[scale=0.25]{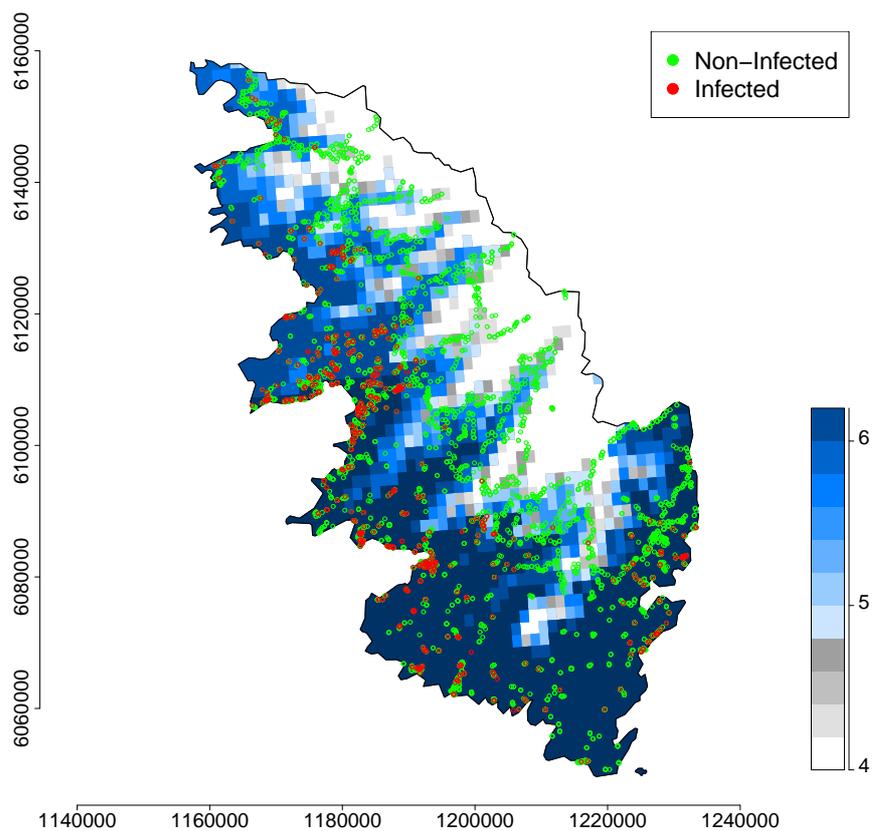}
\caption{Locations of plants, sampled from July 2015 to May 2017, that have been detected as positive (red dots) or negative (green dots) to {\it Xylella fastidiosa} in South Corsica, France, and map of the average of the daily minimum temperature (in Celsius degrees) over January and February reconstructed over a grid with spatial resolution of 1$\times$1km (blue--white color palette).}
\label{fig:1}       
\end{figure}

\begin{figure}[t]
\includegraphics[scale=0.19]{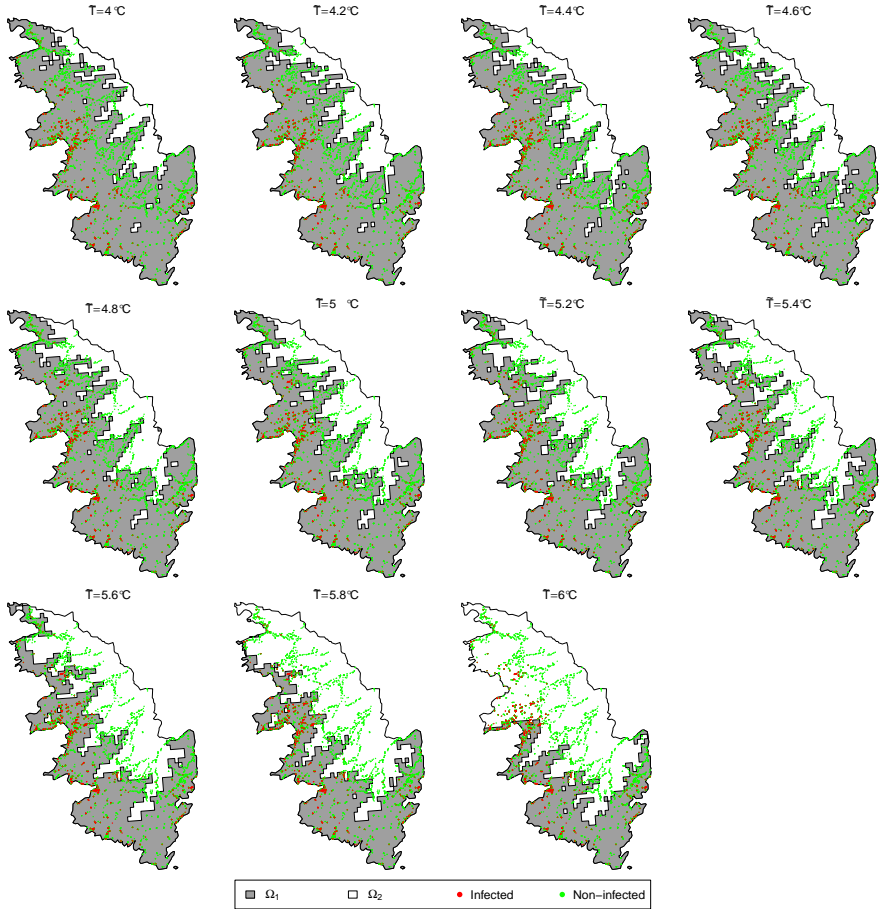}
\caption{Partition of the study domain $\Omega$ into the sub-domains $\Omega_1$ and $\Omega_2$ with respect to the value of $\tilde T$ in $\{4.0,4.2,4.4,\ldots,6.0\}$, in Celsius degrees. Red and green dots give the locations of infected and non-infected samples.}
\label{fig:2}  
\end{figure}

As mentioned in the introduction, we use temperature data to divide the spatial domain into two sub-domains. We exploit a freely available database (PVGIS $\copyright$ European Communities, 2001--2008) providing, in particular, monthly averages of the daily minimum temperature reconstructed over a grid with spatial resolution of 1$\times$1km \citep{huld2006estimating}; these monthly averages correspond to the period 1995-2003, but we used them as references over the period covered by our models.
We use these data to build the average of the daily minimum temperature over January and February, say $T(\mathbf{x})$ for any location $\mathbf{x}$; see Figure \ref{fig:1}. $T(\mathbf{x})$ is then used to split the study domain into two parts: one part where $T(\mathbf{x})\leq\tilde{T}$ and the growth of Xf is hampered by cold winter temperatures, and the other part where $T(\mathbf{x})>\tilde{T}$ and the growth of Xf is not hampered. The threshold value $\tilde{T}$ will be selected in the set $\{4.0,4.2,4.4,\ldots,6.0\}$, in Celsius degrees. Panels of Figure \ref{fig:2} display the partitioning  of the study domain induced by the different values of $\tilde{T}$.

The prior distribution for $\Theta$ combines vague uniform distributions and Dirac distributions:
\begin{equation*}
\begin{split}
[\Theta] =&\dfrac{1}{(10^8-50)\times 100 \times 1 \times 100 \times 1000 \times |\Omega_1|}\\
&\times \mathds{1}(D\in[50;10^8], b\in[0;100], K\in]0;1], \alpha\in[0;100], \tau_0\in[-1000;0], \mathbf{\tilde{x}}_0\in\Omega_1)\\
&\times  \text{Dirac}_{5000}(r_0) \times \text{Dirac}_{0.1}(p_0),
\end{split}
\end{equation*}
where $|\Omega_1|$ is the area of $\Omega_1$ and $\text{Dirac}_b(B)$ is equal to 1 if $B=b$, and 0 otherwise. The Dirac distribution for $\tilde T$ was chosen to deal with implementation issues explained in Section \ref{sec:selection}.
We chose Dirac prior distributions for $r_0$ and $p_0$ in the aim of precisely defining what is an {\it introduction} (see Section \ref{sec:processmodel}) and imposing the same intensity level and spatial extent for the introduction in all the models in competition.
For $D$, $b$, $K$ and $\alpha$, we specified vague uniform priors satisfying constraints of positivity. In addition, the plateau $K$ had to be less than 1, as indicated in Section \ref{sec:processmodel}. 
For the introduction time $\tau_0$, we chose a uniform distribution between $-1000$ months and 0 month before the first detection of Xf in South Corsica. Note that, using a temporal model and aggregated data, \citet{soubeyrand2018inferring} inferred an introduction date around $-360$ months before the first detection of Xf in South Corsica. Finally, the introduction location $\mathbf{\tilde{x}}_0$ was supposed to be uniformly distributed in $\Omega_1$, the sub-domain where the conditions are favorable for the expansion of Xf.

\subsection{Selection of the temperature threshold}
\label{sec:result1}

\begin{figure}[t]
\centering
\includegraphics[scale=0.5]{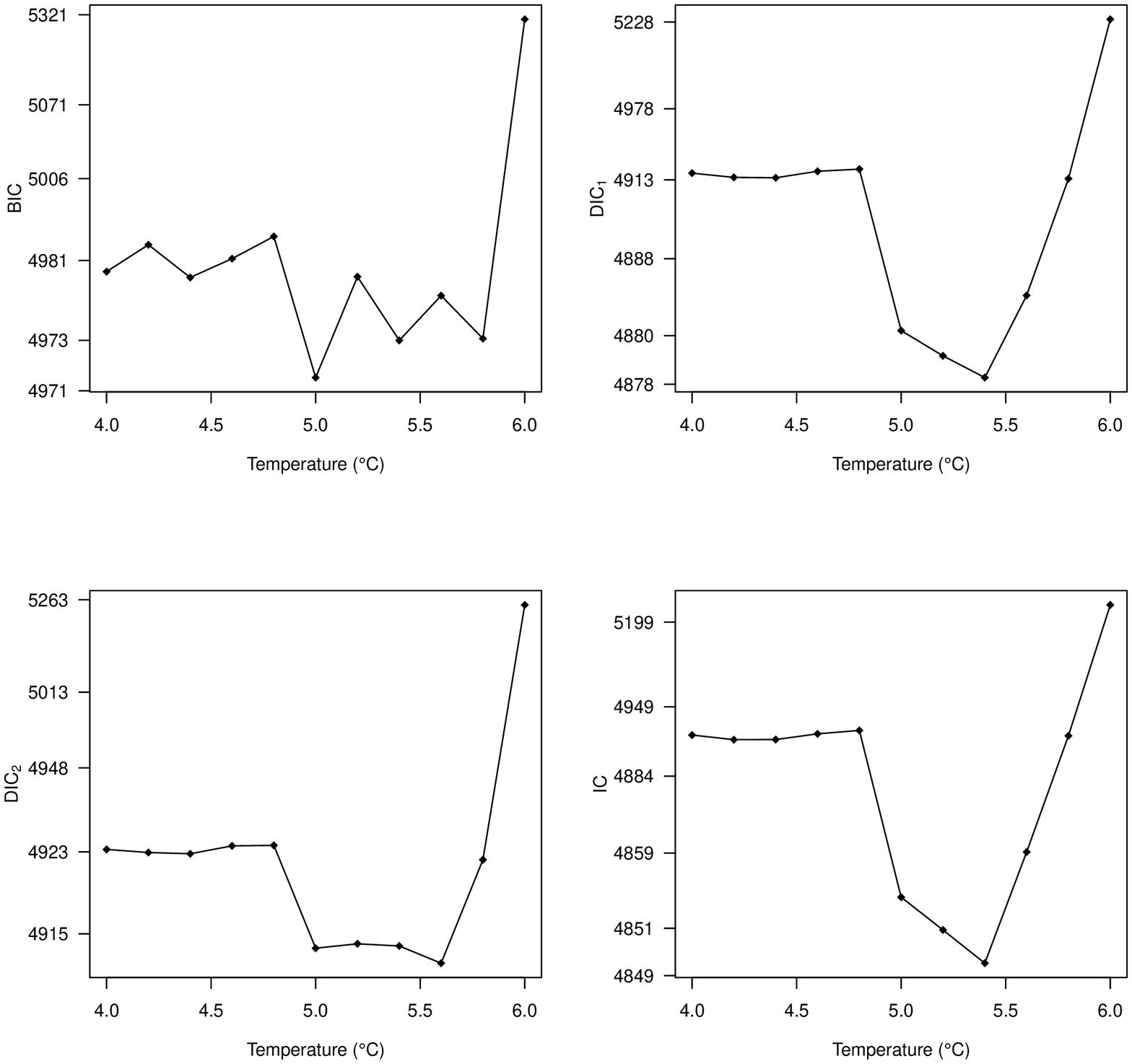}
\caption{Values of the four selection criteria (BIC, $\text{DIC}_1$ of \citet{spiegelhalter2002bayesian}, $\text{DIC}_2$ of \citep{gelman2004bayesian}, IC of \citet{ando2011predictive}) for different  thresholds of temperature  $\tilde{T}$ ranging from 4 to 6$\char23$C. Non-linear transformations of the y-axis were applied to facilitate the identification of the lowest values of the criteria.} 
\label{fig:3}
\end{figure}

The spatio-temporal models corresponding to different values of $\tilde{T}$ ranging from 4 to 6$\char23$C were fitted to data using the estimation approach presented in Section \ref{sec:estimation} (with $(M,L)=(50,10^4)$) and were compared with the four selection criteria introduced in Section \ref{sec:selection}. The values of the criteria are displayed in Figure \ref{fig:3}. The smallest BIC value was obtained for $\tilde{T}=5.0\char23$C. The smallest DIC value based on the computation proposed by \citet{spiegelhalter2002bayesian} and the smallest IC values were obtained for $\tilde{T}=5.4\char23$C. The smallest DIC value based on the computation proposed by \citet{gelman2004bayesian} was obtained for $\tilde{T}=5.6\char23$C. Except the BIC, which only measures the adequacy between the model and data at the posterior mode of the parameter vector, each of the three other criteria takes quite close values around $\tilde{T}=5.4\char23$C (typically from 5.0 to 5.6$\char23$C). In what follows, we present the results obtained with the model corresponding to the threshold $\tilde{T}=5.4\char23$C, which is a satisfying compromise.

\subsection{Stabilization of the AMIS algorithm}\label{sec:stabilization}

The AMIS algorithm provides at each iteration an assessment of the posterior distribution of parameters, which is expected to converge to the true posterior distribution and to be stable after a so-called period of burn-in. If the convergence of AMIS was proved theoretically \citep{cornuet2012adaptive}, we studied its stabilization by evaluating the variation in the following deviation measure between the assessments of the posterior distribution at iteration $m-1$ and $m>1$:
$${\mathcal M}_{\mathcal G}(m-1,m)=\underset{c\in{\mathcal G}}{\max} |p_m(c)-p_{m-1}(c)|,$$
where $p_m(c)$ denotes the assessment at iteration $m$ of the posterior probability that $\Theta$ is in the sub-domain $c\subset {\mathbb R}^8$ of the parameter space, i.e.
$$p_m(c)=\sum_{m'=1}^{m}\sum_{l=1}^{L} w_{m'}^l {\mathds 1}(\Theta_{m'}^l\in c),$$
and $\mathcal G$ is a partition of a sub-space of the parameter space.

 \begin{figure}[t]
 \centering
\begin{turn}{90} 
\hspace*{4cm}\tiny{${\mathcal M}_{\mathcal G}(m-1,m)$ }
\end{turn}\hspace*{0.05cm}    \includegraphics[scale=0.22]{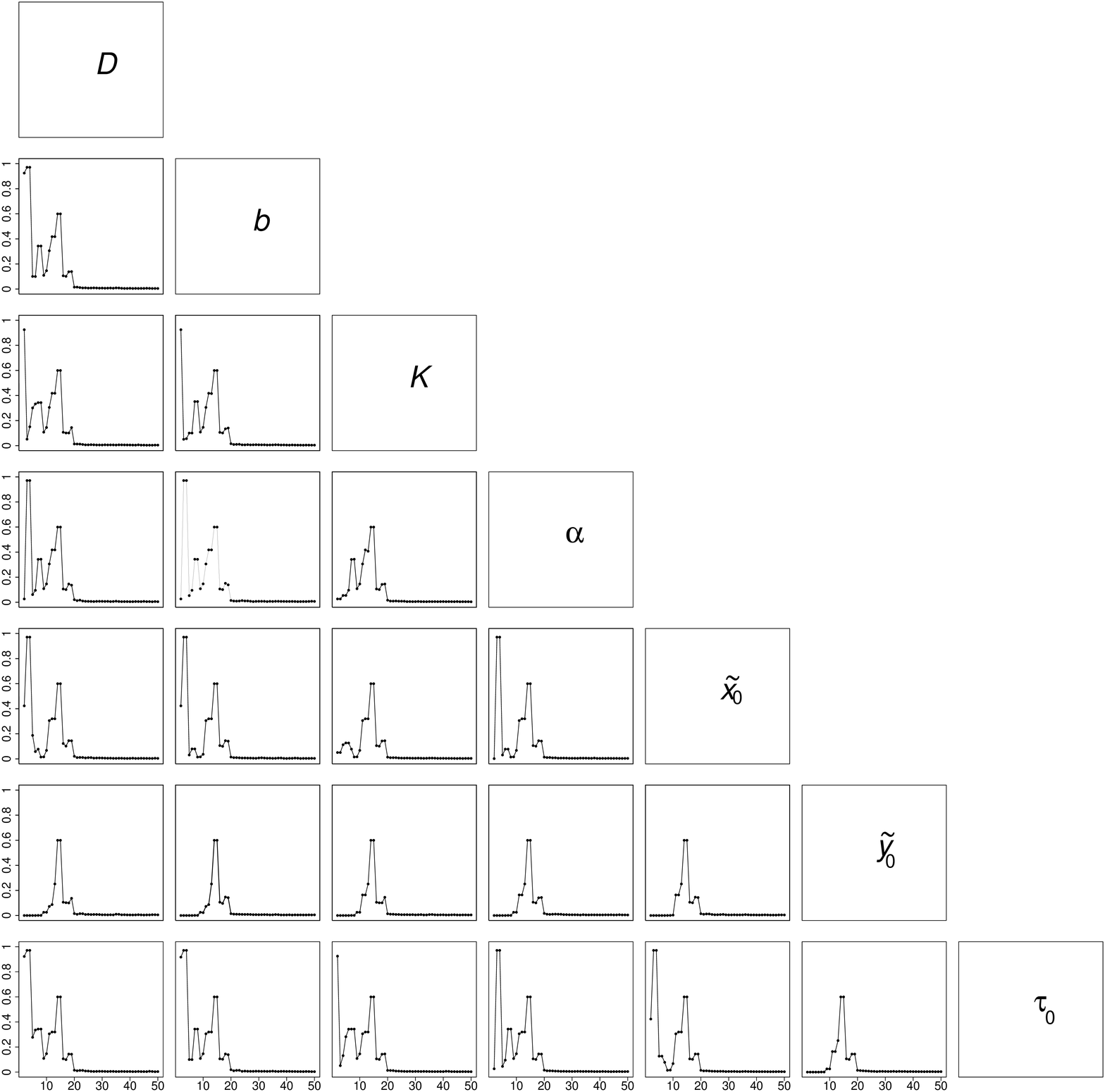}
    \hspace*{1.2cm} \tiny{m} \hspace*{2.3cm} 
    \caption{Variation in the deviation measure ${\mathcal M}_{\mathcal G}(m-1,m)$ between the assessments of the posterior distribution at iteration $m-1$ and $m>1$ of the AMIS algorithm. ${\mathcal M}_{\mathcal G}(m-1,m)$ is plotted for different partitions $\mathcal G$ allowing the assessment of the stabilization of all the 2D posterior distributions of parameters $D$, $b$, $K$, $\alpha$, $\tilde x_0$, $\tilde y_0$ and $\tau_0$.}  
    \label{fig:4}
 \end{figure} 
 
Figure \ref{fig:4} gives the variation in ${\mathcal M}_{\mathcal G}(m-1,m)$ for different partitions $\mathcal G$ allowing us to assess the stabilization of all the 2D posterior distributions of parameters. For each pair of parameters, $\mathcal G$ was defined as the set of infinite cylinders with rectangular bases whose orthogonal projection in the 2 dimensions of interest forms a 60$\times$60 regular rectangular grid. In each dimension of interest, the endpoints of the grid were set at the minimum and maximum values of the corresponding parameter having a weight $w_{M}^l$ larger than $10^{-5}$ (the 2D posterior distributions over these 60$\times$60 grids are displayed in Figure \ref{fig:5}). Figure \ref{fig:4} shows the stabilization of all the 2D posterior distributions after iteration 21.

\subsection{Posterior distribution of parameters}
\label{sec:8}
\begin{figure}[t]
\centering
\includegraphics[height=12cm]{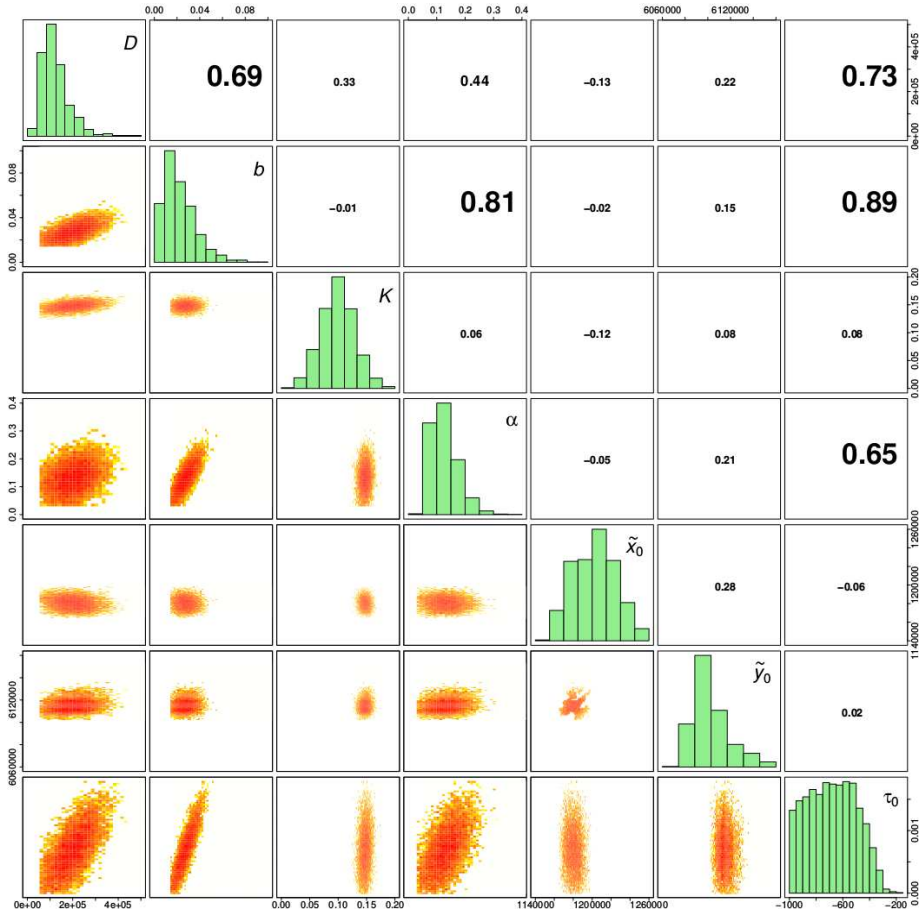}
\caption{Marginal posterior distributions of parameters (panels in the diagonal) and 2D posterior distributions of parameters over the 60$\times$60 grids described in Section \ref{sec:stabilization} (panels in the lower triangle). Figures in the upper triangle panels provide correlation coefficients (the larger the text size, the stronger the correlation).}
\label{fig:5}
\end{figure}
\begin{figure}[t]
\centering
\vspace*{-0.25cm}
\includegraphics[height=7cm]{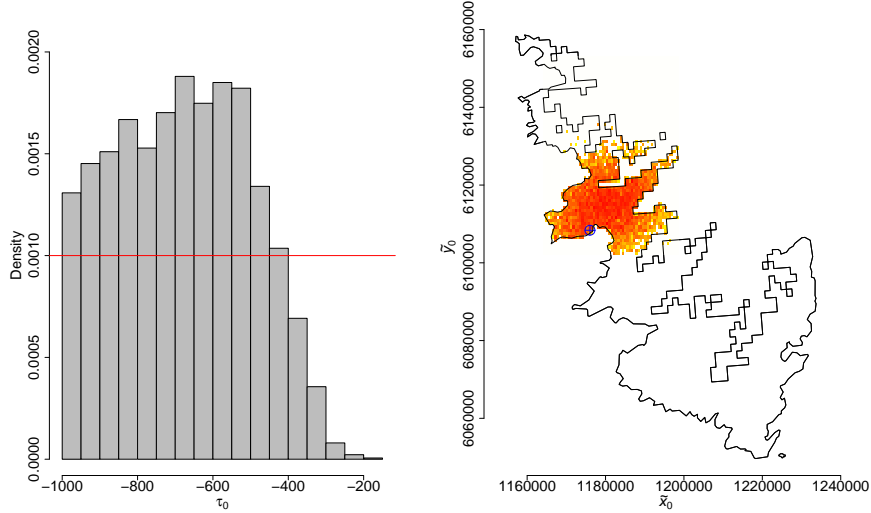}
\caption{Posterior distributions of the introduction time $\tau_0$ (histogram) and the introduction point $\tilde {\bf x}_0$ (color palette). The prior for $\tau_0$ was uniform over $[-1000,0]$ (red line). The value of $\tilde {\bf x}_0$ having the largest weight in AMIS is indicated by a blue dot. The prior for $\tilde {\bf x}_0$ was uniform over the space delimited by the contours.}
\label{fig:6}  
\end{figure}
\begin{figure}
\centering
\includegraphics[height=12cm]{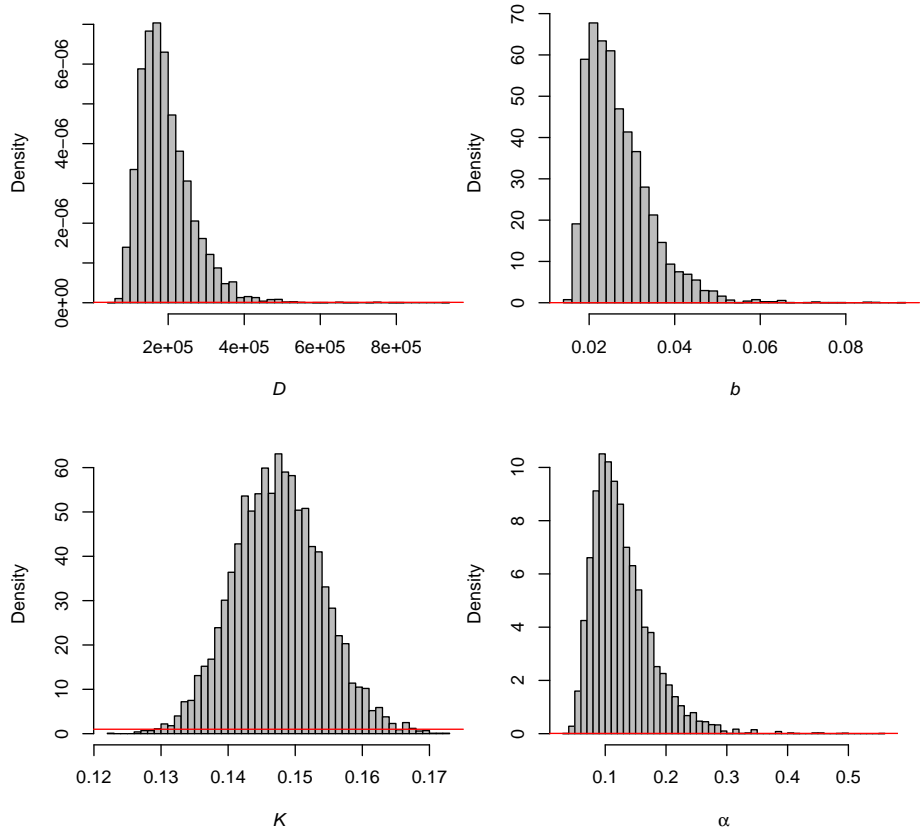}
\caption{Marginal posterior distributions of $D$, $b$, $K$ and $\alpha$ (histograms) and corresponding prior distributions (red lines) over the supports covered by the posteriors.}  
\label{fig:7}
\end{figure}

Marginal and 2D posterior distributions of parameters are displayed in Figures \ref{fig:5}, \ref{fig:6} and \ref{fig:7}. The introduction of {\it Xylella fastidiosa} tends to be relatively ancient (posterior median: $-680$ months before July 2015, i.e. introduction around 1959; posterior mean $-681$ months) but also relatively uncertain (posterior standard deviation: $179$ months). This uncertainty has to be regarded in the light of the relatively high posterior correlation between $\tau_0$ and the reaction-diffusion-absorption parameters $D$, $b$ and $\alpha$. Acquiring knowledge about $D$, $b$ and $\alpha$ could help in eliciting informative priors for these parameters and obtain a less uncertain estimation of the introduction date. Based on our analysis, the introduction probably occurred around Ajaccio or the surrounding municipalities in the East, North and North-East (Right panel of Figure \ref{fig:6}).  Figure \ref{fig:7} and Table \ref{tab:1} show posterior distributions and statistics of $D$, $b$, $K$ and $\alpha$. In particular, we observe that the plateau for the probability of infection is around 15\%. This relatively low estimate has to be considered with caution. First, it is relative to the population of plants that have been sampled. Second, it ignores the risk of false-negative observations. The inference about the diffusion parameter $D$ allowed us to assess the length of a straight line move of the pathogen during a time unit, namely the month. This length is given by Equation \eqref{Dformula} \citep{turchin1998quantitative,roques2016using}:
\begin{equation}
\centering
D =\dfrac{(\text{length of a straight line move during one time step})^2}{4\times \text{duration of the time step}},
\label{Dformula}
\end{equation}
and has a posterior median equal to 155 meters per month (posterior mean: 155; posterior standard deviation: 27). These figures correspond to the move of the pathogen with different means, in particular via insects and transportation of infected plants, which are both modeled by the diffusion operator in Equation \eqref{xf}.


\begin{table}[t]
\centering
\caption{Posterior medians, means and standard deviations of parameters of the reaction-diffusion-absorption equation.}
\label{tab:1}      
\begin{tabular}{llccc}
\hline\noalign{\smallskip}
 Parameter & Unit &Median & Mean & Standard deviation   \\
\noalign{\smallskip}\hline\noalign{\smallskip}
$D$ & m$^2\cdot$ month$^{-1}$ & $1.8\times 10^{5}$  & $2.0\times 10^{5}$  & $0.7\times 10^{5}$  \\
$b$ & month$^{-1}$ & $0.026$ & $0.027$ & $0.008$ \\
$K$ & probability & $0.147$ & $0.148$ & $0.007$ \\
$\alpha$ & month$^{-1}$ & $0.12$  & $0.13$  & $0.05$  \\
\noalign{\smallskip}\hline
\end{tabular}
\end{table}

\subsubsection{Goodness-of-fit of the model}
\label{sec:9}

To check the adequacy between the selected model and observed data, we measured the accuracy of the probabilistic predictions provided by the model by using the Brier score (BS) \citep{brier1950verification}. This score is the mean of the square differences between (i) the observed health statuses $Y_i$ (in $\{0,1\}$), $i=1,\ldots,I$, and (ii) the corresponding probabilities of infection $u(t_i,\mathbf{x}_i)$, which depend on~$\Theta$:
\begin{equation}
\centering
\text{BS}=\dfrac{1}{I}\sum_{i=1}^I\big(Y_i-u(t_i,\mathbf{x}_i)\big)^2.
\end{equation}
The Brier score varies between 0 and 1; lower the Brier score, better the goodness-of-fit; a systematic prediction of 0.5 leads to a Brier score equal to 0.25, which can be viewed as a threshold above which the model is clearly inadequate. 
In our application, the posterior median of BS is 0.0829 (95\%-posterior interval: [0.0827,0.0830]). 

\begin{figure}[t]
\centering
\includegraphics[height=7.5cm]{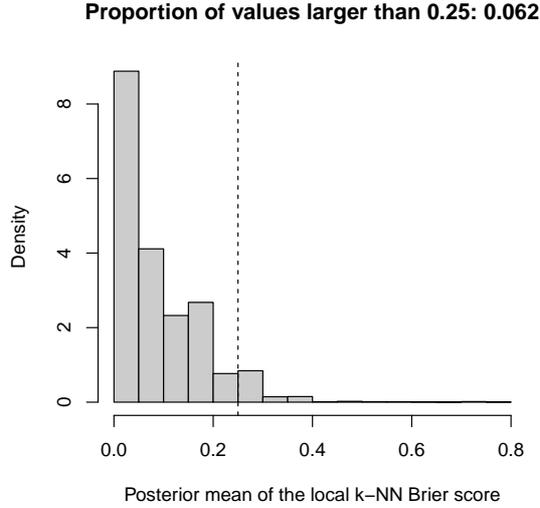}
\caption{Distribution of the posterior means of the local Brier scores. The dashed line gives the 0.25 threshold.}
\label{fig:8}
\end{figure}

\begin{figure}
\centering
\vspace*{-1.3cm}
\includegraphics[scale=0.4]{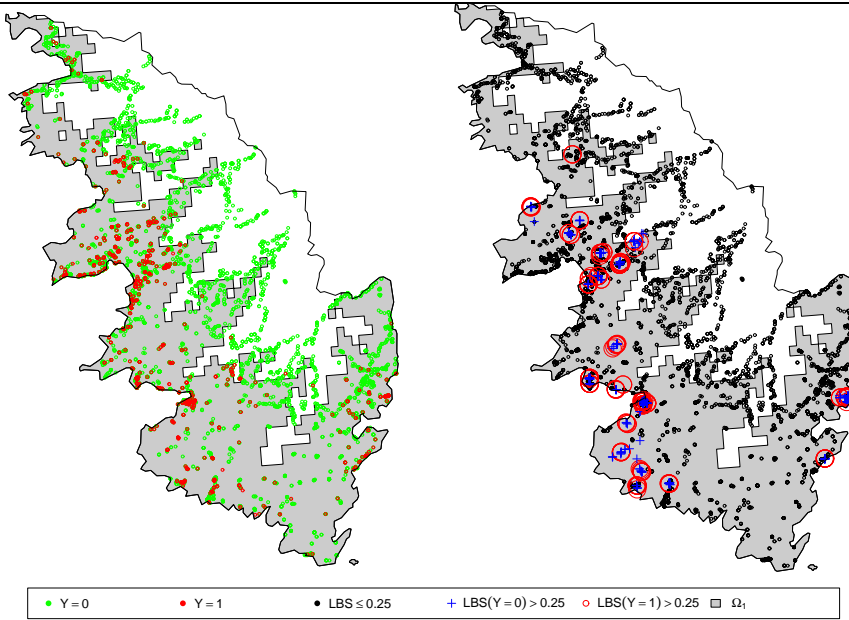}
\vspace*{-1.3cm}
\caption{Locations of samples diagnosed as positive and negative to {\it Xylella fastidiosa} (left) and samples with different levels of the local Brier score (right; black dots: LBS$_k(i)\le 0.25$; blue crosses: $Y_i=0$ and LBS$_k(i)> 0.25$; red circles: $Y_i=1$ and LBS$_k(i)> 0.25$). The gray surface gives the extent of $\Omega_1$.}
\label{fig:9}
\end{figure}

We extended the goodness-of-fit analysis by building and analyzing a local Brier score that allows us to check the adequacy of the model across space. The local Brier score (LBS) computed at the location of observation $i\in\{1,\ldots,I\}$ is defined as a mean over the $k$-nearest neighbors:
\begin{equation}
\centering
\text{LBS}_k(i)=\frac{1}{k+1}\sum_{i'\in\{i\}\cup\mathcal{V}_k(i)}\big(Y_{i'}-u(t_{i'},\mathbf{x}_{i'})\big)^2,
\end{equation}
where $\mathcal{V}_k(i)$ is the set of indices in $\{1,\ldots,I\}$ corresponding to the $k>0$ observations nearest to $\mathbf{x}_i$ with respect to the Euclidean distance in $\mathds R^2$. 
Figure \ref{fig:8} gives the distribution of the posterior means of the local Brier scores. 6.2\% of these scores are above 0.25, which is a rather small percentage.
Figure \ref{fig:9} displays locations where the LBS is larger than 0.25 with $k=20$ (Supplementary Figure S3 provides similar information for $k$ equal to 50, 100 and 150). This figure also indicates whether observations with LBS$>$0.25 were detected as positive or negative to {\it Xylella fastidiosa}. None of the observations with LBS$>$0.25 are in $\Omega_2$ where the growth of the pathogen is negative. Thus, discrepancies between data and the model are limited to $\Omega_1$. In addition, in general, model discrepancies for positive samples and negative samples are located approximately at the same places. Therefore, there might be some spatially abrupt changes in the rate of infection that are not represented by our aggregated model, which is based on a reaction-diffusion-absorption equation mostly adapted to represent smooth variations in space.

\section{Discussion}
\label{sec:discussion}

Since the detection of \textit{Xylella fastidiosa} in Europe, several modeling approaches have been implemented to provide more insights on the spread of this invasive pathogen in European environments \citep{strona2017network,white_modelling_2017,bosso_potential_2016,godefroid2018climate,soubeyrand2018inferring,martinetti2018}. In this paper, we mainly focus on dating and localizing the introduction of this invasive species. Nevertheless, inferring the parameters of the coupled reaction-diffusion-absorption equation is required since only post-introduction data are available. The conducted analyses using a Bayesian inference approach, tend to show that the introduction of \textit{Xylella fastidiosa} in South Corsica occurred probably near Ajaccio around 1959 (95\%-posterior interval: $[1933,1986]$), long time before its first detection.
Our estimation of the introduction time is relatively consistent with the results obtained by \citet{denanceisxylella} who assessed the introduction of the two main strains found in Corsica around 1965 and 1980, respectively, using a phylogenetic approach. Likewise, our estimation is compatible to the result of \citet{soubeyrand2018inferring}, who dated the introduction around 1985 (95\%-posterior interval: $[1978,1993]$) with a statistical analysis of temporal data (indeed, the posterior intervals obtained from both analyses overlap). To obtain a more accurate estimation of the introduction date, at least two tracks could be followed: coupling the analysis of spatio-temporal surveillance data and genetic data, as discussed in \citet{soubeyrand2018inferring}, and, as suggested in the result section, gaining knowledge about parameters $D$, $b$ and $\alpha$ whose estimations are correlated with the estimation of the introduction date (such a knowledge could be incorporated into the prior distribution and could lead to a narrower posterior distribution of $\tau_0$).

To infer the posterior distribution of the parameter vector we proceed in two steps: (i) infer the parameters of the dynamics given the temperature threshold $\tilde{T}$ used for partitioning the study domain, and (ii) choose $\tilde{T}$ using different selection criteria. A possible extension of our work is to refine the definition of the spatial partition by not only using the minimum daily winter temperature but also other relevant environmental variables \citep{godefroid2018climate,martinetti2018}. Thus, a parametric logistic regression function depending on these variables could be built for partitioning the study domain and its parameters should be jointly estimated with the other parameters. However, this perspective requires a faster estimation approach. Indeed, an important milestone towards an accurate inference about the parameter vector, is to accurately solve the partial differential equation, which requires non-negligible computation time. Fortunately, the AMIS algorithm is easily parallelized (with $(M,L)=(50,10^4)$ and the use of $100$ cluster cores, the estimation procedure for a fixed value of $\tilde{T}$ took approximately $1.75$ days). However, jointly estimating the partition of the study domain (and not only selecting it as we did), would result on much larger computation times, especially if the partition depends on multiple spatial variables. To reduce the computational cost, approximating the input/output relation in the mechanistic model using meta-models necessiting less computer intensive calculations could be a valuable option, that could be incorporated in AMIS \citep{osio1996engineering,giunta1998comparison}. In particular, kriging meta-models show up to be an adequate solution for approximating deterministic models since they interpolate the observed or known data points \citep{simpson2001metamodels}. An additional advantage that derives from the use of AMIS is that its tuning parameters are adapted across the algorithm iterations differently from the basic MCMC, maximum of likelihood and other frequently used approaches in the mechanistic-statistical framework. It has however to be noted that AMIS has to be appropriately initialized, which can be relatively easily done in practice by evaluating the marginal posterior distributions over 1D grids.

Obviously, the deterministic model (Equation (\ref{xf}-\ref{initialcondition})) that we proposed to describe the dynamics of the pathogen does not take into account all the epidemiological and environmental drivers of the dynamics. These drivers could be implicitly handled by replacing our model by a stochastic version that would result in more flexible realizations. \citet{gonze2002deterministic} compared  deterministic  and  stochastic  models  for circadian  oscillations and showed that, in presence of noise in a small population, stochastic simulations are needed to get more realistic realizations. Although the population size for the case study of \textit{Xylella fastidiosa} is expected to be relatively large, stochastic population-dynamic models, from individual-based models \citep{renshaw1993modelling,kareiva1983analyzing} to aggregated models \citep{soubeyrand_2009b}, could allow to relax hypotheses made on the dynamics. Nevertheless, a stochastic version of our model, especially if it incorporates latent variables to be estimated, will induce an additional (heavy) computational cost in the inference procedure. In contrast, our parsimonious model, which only incorporates the main epidemiological and environmental drivers, provides a concise description of the dynamics of the pathogen, and can be fitted to data in a reasonable time span. The advantage of this approach is that it can be rapidly applied for endorsing a fast reaction after the detection of a new invasive pathogen. An intermediate modeling approach for gaining in realism but conserving parsimony, could be to couple the conciseness of our reaction-diffusion-absorption equation with a stochastic jump process in the settings of Piecewise Deterministic Markov Process \citep[PDMP;][]{azais2018}. In this framework, each jump could be localized in space and time and would represent a new introduction of the pathogen, as proposed by \citet{abboud2018pdmp}. A strong limitation of the number of jumps would be needed to keep the model and the estimation of its parameters simple. Such a model would be relevant for our application since the emergence of \textit{Xylella fastidiosa} in France is likely to be linked to the introduction of several pathogen strains \citep{denance2017several}.

\bibliographystyle{spbasic}      
\bibliography{article.bbl}   

\begin{thebibliography}{75}
\providecommand{\natexlab}[1]{#1}
\providecommand{\url}[1]{{#1}}
\providecommand{\urlprefix}{URL }
\expandafter\ifx\csname urlstyle\endcsname\relax
  \providecommand{\doi}[1]{DOI~\discretionary{}{}{}#1}\else
  \providecommand{\doi}{DOI~\discretionary{}{}{}\begingroup
  \urlstyle{rm}\Url}\fi
\providecommand{\eprint}[2][]{\url{#2}}

\bibitem[{Abboud et~al(2018)Abboud, Senoussi, and Soubeyrand}]{abboud2018pdmp}
Abboud C, Senoussi R, Soubeyrand S (2018) {Piecewise-deterministic Markov
  Processes for Spatio-temporal Population Dynamics}. In: {Aza{\"\i}s, Romain
  and Bouguet, Florian} (ed) {Statistical Inference for Piecewise-deterministic
  Markov Processes}, ISTE Editions/Wiley

\bibitem[{Allaire(2008)}]{ALL2008}
Allaire G (2008) Analyse num\'erique et optimisation. Les \'Editions de
  l'\'Ecole Polytechnique, Palaiseau

\bibitem[{Anas et~al(2008)Anas, Harrison, Brannen, and Sutton}]{anas2008effect}
Anas O, Harrison UJ, Brannen PM, Sutton TB (2008) The effect of warming winter
  temperature on the severity of pierce’s disease in the appalachian
  mountains and piedmont of the southeastern united states. Plant Health
  Progress 101094:450--459

\bibitem[{Anderson et~al(1996)Anderson, Donnelly, Ferguson, Woolhouse, Watt,
  Udy, Mawhinney, Dunstan, Southwood, Wilesmith, Ryan, Hoinville, Hillerton,
  Austin, and Wells}]{anderson_transmission_1996}
Anderson RM, Donnelly CA, Ferguson NM, Woolhouse MEJ, Watt CJ, Udy HJ,
  Mawhinney S, Dunstan SP, Southwood TRE, Wilesmith JW, Ryan JBM, Hoinville LJ,
  Hillerton JE, Austin AR, Wells GAH (1996) Transmission dynamics and
  epidemiology of {BSE} in {British} cattle. Nature 382:779--788,
  \urlprefix\url{https://doi.org/10.1038/382779a0}

\bibitem[{Ando(2011)}]{ando2011predictive}
Ando T (2011) Predictive {B}ayesian model selection. American Journal of
  Mathematical and Management Sciences 31:13--38,
  \urlprefix\url{https://doi.org/10.1080/01966324.2011.10737798}

\bibitem[{Andow et~al(1990)Andow, Kareiva, Levin, and Okubo}]{andow1990spread}
Andow D, Kareiva PM, Levin SA, Okubo A (1990) Spread of invading organisms.
  Landscape Ecology 4:177--188

\bibitem[{Andow et~al(1993)Andow, Kareiva, Levin, and Okubo}]{andow1993spread}
Andow D, Kareiva P, Levin S, Okubo A (1993) Spread of invading organisms:
  patterns of spread. Evolution of insect pests: the pattern of variations pp
  219--242

\bibitem[{Aza{\"\i}s and Bouguet(2018)}]{azais2018}
Aza{\"\i}s R, Bouguet F (2018) Statistical Inference for
  Piecewise-deterministic Markov Processes. ISTE Editions/Wiley

\bibitem[{Baker(1991)}]{baker1991continuing}
Baker HG (1991) The continuing evolution of weeds. Economic Botany 45:445--449

\bibitem[{Berliner(2003)}]{berliner2003physical}
Berliner LM (2003) Physical-statistical modeling in geophysics. Journal of
  Geophysical Research: Atmospheres 108:8776,
  \urlprefix\url{https://doi.org/10.1029/2002JD002865}

\bibitem[{Bosso et~al(2016)Bosso, Russo, Febbraro, Cristinzio, and
  Zoina}]{bosso_potential_2016}
Bosso L, Russo D, Febbraro MD, Cristinzio G, Zoina A (2016) {Potential
  distribution of {\it Xylella fastidiosa} in {Italy}: a maximum entropy
  model}. Phytopathologia Mediterranea 55:62--72

\bibitem[{Boys et~al(2008)Boys, Wilkinson, and Kirkwood}]{boys2008bayesian}
Boys RJ, Wilkinson DJ, Kirkwood TBL (2008) Bayesian inference for a discretely
  observed stochastic kinetic model. Statistics and Computing 18:125--135,
  \urlprefix\url{https://doi.org/10.1007/s11222-007-9043-x}

\bibitem[{Brier(1950)}]{brier1950verification}
Brier GW (1950) Verification of forecasts expressed in terms of probability.
  OPTmonthey Weather Review 78:1--3

\bibitem[{Brooks(2003)}]{Brooks03}
Brooks S (2003) {B}ayesian computation~: A statistical revolution. Trans Roy
  Statist Soc, series A 15:2681--2697,
  \urlprefix\url{{https://doi.org/10.1098/rsta.2003.1263}}

\bibitem[{Bugallo et~al(2015)Bugallo, Martino, and
  Corander}]{bugallo2015adaptive}
Bugallo MF, Martino L, Corander J (2015) Adaptive importance sampling in signal
  processing. Digital Signal Processing 47:36--49,
  \urlprefix\url{https://doi.org/10.1016/j.dsp.2015.05.014}

\bibitem[{Chapman et~al(2015)Chapman, White, Hooftman, and
  Bullock}]{chapman2015inventory}
Chapman DS, White SM, Hooftman DA, Bullock JM (2015) {Inventory and review of
  quantitative models for spread of plant pests for use in pest risk assessment
  for the EU territory}. EFSA Supporting Publications 12,
  \urlprefix\url{https://doi.org/10.2903/sp.efsa.2015.EN-795}

\bibitem[{Cornuet et~al(2012)Cornuet, Marin, Mira, and
  Robert}]{cornuet2012adaptive}
Cornuet J, Marin JM, Mira A, Robert CP (2012) Adaptive multiple importance
  sampling. Scandinavian Journal of Statistics 39:798--812,
  \urlprefix\url{https://doi.org/10.1111/j.1467-9469.2011.00756.x}

\bibitem[{Costello et~al(2017)Costello, Steinmaus, and
  Boisseranc}]{costello_environmental_2017}
Costello M, Steinmaus S, Boisseranc C (2017) Environmental variables
  influencing the incidence of {Pierce}'s disease. Australian Journal of Grape
  and Wine Research 23:287--295,
  \urlprefix\url{https://doi.org/10.1111/ajgw.12262}

\bibitem[{Denanc{\'e} et~al(2017{\natexlab{a}})Denanc{\'e}, Cesbron, Briand,
  Rieux, and Jacques}]{denanceisxylella}
Denanc{\'e} N, Cesbron S, Briand M, Rieux A, Jacques MA (2017{\natexlab{a}})
  {Is {\it Xylella fastidiosa} really emerging in {F}rance? }. In: Costa J,
  Koebnik R (eds) {1st Annual Conference of the EuroXanth - {COST A}ction
  {I}ntegrating {S}cience on {\it Xanthomonadaceae} for integrated plant
  disease management in {E}urope}, {E}uro{X}anth, {Coimbra, Portugal}, vol~7

\bibitem[{Denanc{\'e} et~al(2017{\natexlab{b}})Denanc{\'e}, Legendre, Briand,
  Olivier, Boisseson, Poliakoff, and Jacques}]{denance2017several}
Denanc{\'e} N, Legendre B, Briand M, Olivier V, Boisseson C, Poliakoff F,
  Jacques MA (2017{\natexlab{b}}) {Several subspecies and sequence types are
  associated with the emergence of {\it Xylella fastidiosa} in natural settings
  in France}. {Plant Pathology} 66:1054--1064,
  \urlprefix\url{https://doi.org/10.1111/ppa.12695}

\bibitem[{Evans(1998)}]{evans_partial_1998}
Evans LC (1998) Partial differential equations, Graduate studies in
  mathematics, vol~19. American Mathematical Society, Providence, Rhode Island

\bibitem[{Feil and Purcell(2001)}]{feil_temperature-dependent_2001}
Feil H, Purcell AH (2001) {Temperature-dependent growth and survival of {\it
  {Xylella} fastidiosa} in vitro and in potted grapevines}. Plant Disease
  85:1230--1234, \urlprefix\url{https://doi.org/10.1094/PDIS.2001.85.12.1230}

\bibitem[{Feil et~al(2003)Feil, Feil, and Purcell}]{feil_effects_2003}
Feil H, Feil WS, Purcell AH (2003) {Effects of date of inoculation on the
  within-plant movement of {\it {Xylella} fastidiosa} and persistence of
  {Pierce}'s disease within field grapevines}. Phytopathology 93:244--251,
  \urlprefix\url{http://dx.doi.org/10.1094/PHYTO.2003.93.2.244}

\bibitem[{Fisher(1937)}]{fisher_wave_1937}
Fisher RA (1937) {The} {wave} {of} {advance} {of} {advantageous} {genes}.
  Annals of Eugenics 7:355--369,
  \urlprefix\url{https://doi.org/10.1111/j.1469-1809.1937.tb02153.x}

\bibitem[{Gatenby and Gawlinski(1996)}]{gatenby_reaction-diffusion_1996}
Gatenby RA, Gawlinski ET (1996) A reaction-diffusion model of cancer invasion.
  Cancer research 56:5745--5753

\bibitem[{Gelfand and Smith(1990)}]{gelfand1990sampling}
Gelfand AE, Smith AFM (1990) {Sampling-Based Approaches to Calculating Marginal
  Densities}. Journal of the American Statistical Association 85:398--409,
  \urlprefix\url{https://doi.org/10.1080/01621459.1990.10476213}

\bibitem[{Gelman et~al(1996)Gelman, Roberts, Gilks et~al}]{gelman1996efficient}
Gelman A, Roberts GO, Gilks WR, et~al (1996) Efficient metropolis jumping
  rules. {B}ayesian statistics 5:599--608

\bibitem[{Gelman et~al(2003)Gelman, Carlin, Stern, and
  Rubin}]{gelman2004bayesian}
Gelman A, Carlin JB, Stern HS, Rubin DB (2003) {{B}ayesian data analysis}, 2nd
  edn. {Texts in statistical science series}, Chapman \& Hall/CRC, New York

\bibitem[{Giunta and Watson(1998)}]{giunta1998comparison}
Giunta A, Watson L (1998) {A comparison of approximation modeling
  techniques-Polynomial versus interpolating models}. In: {7th
  AIAA/USAF/NASA/ISSMO Symposium on Multidisciplinary Analysis and
  Optimization}, Multidisciplinary Analysis Optimization Conferences, {St.
  Louis, MO, U.S.A.}, p 4758, \urlprefix\url{https://doi.org/10.2514/MMAO98}

\bibitem[{Godefroid et~al(2018)Godefroid, Cruaud, Streito, Rasplus, and
  Rossi}]{godefroid2018climate}
Godefroid M, Cruaud A, Streito JC, Rasplus JY, Rossi JP (2018) {Climate change
  and the potential distribution of Xylella fastidiosa in Europe}. bioRxiv
  \doi{10.1101/289876}

\bibitem[{Gonze et~al(2002)Gonze, Halloy, and
  Goldbeter}]{gonze2002deterministic}
Gonze D, Halloy J, Goldbeter A (2002) {Deterministic versus stochastic models
  for circadian rhythms}. Journal of biological physics 28:637--653,
  \urlprefix\url{{https://doi.org/10.1023/A:1021286607354}}

\bibitem[{Hecht(2012)}]{hecht2012new}
Hecht F (2012) New development in {F}reefem++. Journal of numerical mathematics
  20:251--266, \urlprefix\url{https://doi.org/10.1515/jnum-2012-0013}

\bibitem[{Hengeveld(1989)}]{hengeveld_dynamics_1989}
Hengeveld R (1989) Dynamics of biological invasions. Springer Science \&
  Business Media, New York

\bibitem[{Henneberger(2003)}]{henneberger2003effects}
Henneberger TS (2003) {Effects of low temperature on populations of {\it
  {X}ylella fastidiosa} in sycamore}. PhD thesis, University of Georgia

\bibitem[{Huld et~al(2006)Huld, {\v{S}}{\'u}ri, Dunlop, and
  Micale}]{huld2006estimating}
Huld TA, {\v{S}}{\'u}ri M, Dunlop ED, Micale F (2006) Estimating average
  daytime and daily temperature profiles within europe. Environmental Modelling
  \& Software 21:1650--1661

\bibitem[{Kareiva and Shigesada(1983)}]{kareiva1983analyzing}
Kareiva P, Shigesada N (1983) {Analyzing insect movement as a correlated random
  walk}. {Oecologia} 56:234--238,
  \urlprefix\url{{https://doi.org/10.1007/BF00379695}}

\bibitem[{Kermack and McKendrick(1927)}]{kermack_contributions_1991}
Kermack WO, McKendrick AG (1927) A contribution to the mathematical theory of
  epidemics. The Royal Society 115:700--721,
  \urlprefix\url{https://doi.org/10.1098/rspa.1927.0118}

\bibitem[{Lanzarone et~al(2017)Lanzarone, Pasquali, Gilioli, and
  Marchesini}]{lanzarone2017bayesian}
Lanzarone E, Pasquali S, Gilioli G, Marchesini E (2017) A {B}ayesian estimation
  approach for the mortality in a stage-structured demographic model. Journal
  of mathematical biology 75:759--779,
  \urlprefix\url{https://doi.org/10.1007/s00285-017-1099-4}

\bibitem[{{Lewis, MA and Kareiva, P}(1993)}]{lewis1993}
{Lewis, MA and Kareiva, P} (1993) {Allee dynamics and the spread of invading
  organisms}. Theoretical Population Biology 43:141--158,
  \urlprefix\url{https://doi.org/10.1006/tpbi.1993.1007}

\bibitem[{Lindley(2006)}]{Lin2006}
Lindley D (2006) {Understanding Uncertainty}. John Wiley \& Sons, INC, Hoboken,
  New Jersey, \urlprefix\url{https://doi.org/10.1002/0470055480}

\bibitem[{Martinetti and Soubeyrand(2018)}]{martinetti2018}
Martinetti D, Soubeyrand S (2018) {Identifying lookouts for
  epidemio-surveillance: application to the emergence of {\it Xylella
  fastidiosa} in {F}rance}, submitted

\bibitem[{Mollison(1977)}]{mollison_spatial_1977}
Mollison D (1977) Spatial contact models for ecological and epidemic spread.
  Journal of the Royal Statistical Society Series B (Methodological)
  39:283--326

\bibitem[{Murray(2002)}]{murray2002mathematical}
Murray JD (2002) Mathematical biology. In: Interdisciplinary Applied
  Mathematics, vol~17, 3rd edn, Springer-Verlag, New York

\bibitem[{Okubo(1980)}]{okubo_diffusion_1980}
Okubo A (1980) Diffusion and ecological problems: mathematical models,
  Interdisciplinary Applied Mathematics, vol~10. Springer-Verlag, New York

\bibitem[{Okubo and Levin(2002)}]{okubo2002diffusion}
Okubo A, Levin S (2002) Diffusion and ecological problems - {M}odern
  {Perspectives}, 2nd edn. New York,
  \urlprefix\url{https://doi.org/10.1007/978-1-4757-4978-6}

\bibitem[{Osio and Amon(1996)}]{osio1996engineering}
Osio IG, Amon CH (1996) An engineering design methodology with multistage
  bayesian surrogates and optimal sampling. {Research in Engineering Design}
  8:189--206

\bibitem[{Peterson et~al(2003)Peterson, Vucetich, Page, Chouinard
  et~al}]{peterson2003temporal}
Peterson RO, Vucetich JA, Page RE, Chouinard A, et~al (2003) {Temporal and
  spatial aspects of predator--prey dynamics}. Alces 39:215--232,
  \urlprefix\url{https://doi.org/10.1098/rspb.2015.0973}

\bibitem[{{Protter, {MH} and Weinberger, {HF}}(1967)}]{protter1967maximum}
{Protter, {MH} and Weinberger, {HF}} (1967) Maximum Principles in Differential
  Equations. {Prentice-Hall, Englewood Cliffs}, New Jersey,
  \urlprefix\url{https://doi.org/10.1007/978-1-4612-5282-5}

\bibitem[{Purcell(1977)}]{purcell1977cold}
Purcell A (1977) Cold therapy of pierce's disease of grapevines. Plant Disease
  Reporter 61:514--518

\bibitem[{Purcell et~al(1980)}]{purcell1980environmental}
Purcell A, et~al (1980) Environmental therapy for pierce's disease of
  grapevines. Plant Disease 64:388--390

\bibitem[{Reise et~al(2006)Reise, Olenin, and Thieltges}]{reise2006aliens}
Reise K, Olenin S, Thieltges DW (2006) Are aliens threatening european aquatic
  coastal ecosystems? Helgoland Marine Research 60:77,
  \urlprefix\url{https://doi.org/10.1007/s10152-006-0024-9}

\bibitem[{Renshaw(1993)}]{renshaw1993modelling}
Renshaw E (1993) {Modelling biological populations in space and time}, vol~11.
  Cambridge University Press, Cambridge,
  \urlprefix\url{https://doi.org/10.1017/CBO9780511624094}

\bibitem[{Richardson and Bond(1991)}]{richardson1991determinants}
Richardson DM, Bond WJ (1991) Determinants of plant distribution: evidence from
  pine invasions. The American Naturalist 137:639--668

\bibitem[{Roberts et~al(1997)Roberts, Gelman, and Gilks}]{roberts1997weak}
Roberts GO, Gelman A, Gilks WR (1997) Weak convergence and optimal scaling of
  random walk metropolis algorithms. The Annals of Applied Probability
  7:110--120

\bibitem[{Roques et~al(2011)Roques, Soubeyrand, and Rousselet}]{roques2011}
Roques L, Soubeyrand S, Rousselet J (2011) A statistical-reaction--diffusion
  approach for analyzing expansion processes. Journal of Theoretical Biology
  274:43--51, \urlprefix\url{https://doi.org/10.1016/j.jtbi.2011.01.006}

\bibitem[{Roques et~al(2016)Roques, Walker, Franck, Soubeyrand, and
  Klein}]{roques2016using}
Roques L, Walker E, Franck P, Soubeyrand S, Klein E (2016) Using genetic data
  to estimate diffusion rates in heterogeneous landscapes. Journal of
  mathematical biology 73:397--422,
  \urlprefix\url{https://doi.org/10.1007/s00285-015-0954-4}

\bibitem[{Schwarz et~al(1978)}]{schwarz1978estimating}
Schwarz G, et~al (1978) Estimating the dimension of a model. The annals of
  statistics 6:461--464, \urlprefix\url{https://doi.org/10.1214/aos/1176344136}

\bibitem[{Shigesada and Kawasaki(1997)}]{shigesada_biological_1997}
Shigesada N, Kawasaki K (1997) Biological invasions: theory and practice, 1st
  edn. Oxford series in ecology and evolution, Oxford University Press, Oxford,
  New York

\bibitem[{Shigesada et~al(1995)Shigesada, Kawasaki, and
  Takeda}]{shigesada_modeling_1995}
Shigesada N, Kawasaki K, Takeda Y (1995) Modeling {Stratified} {Diffusion} in
  {Biological} {Invasions}. The American Naturalist 146:229--251

\bibitem[{Simberloff(1989)}]{simberloff1989insect}
Simberloff D (1989) Which insect introductions succeed and which fail?, vol~37,
  Wiley, Chichester, UK, pp 61--75

\bibitem[{Simpson et~al(2001)Simpson, Poplinski, Koch, and
  Allen}]{simpson2001metamodels}
Simpson TW, Poplinski J, Koch PN, Allen JK (2001) {Metamodels for
  computer-based engineering design: survey and recommendations}. {Engineering
  with computers} 17:129--150,
  \urlprefix\url{{https://doi.org/10.1007/PL00007198}}

\bibitem[{Skellam(1951)}]{skellam_random_1951}
Skellam JG (1951) Random dispersal in theoretical populations. Biometrika
  38:196--218, \urlprefix\url{https://doi.org/10.2307/2332328}

\bibitem[{Soubeyrand and Roques(2014)}]{soubeyrand_parameter_2014}
Soubeyrand S, Roques L (2014) Parameter estimation for reaction-diffusion
  models of biological invasions. Population Ecology 56:427--434,
  \urlprefix\url{https://doi.org/10.1007/s10144-013-0415-0}

\bibitem[{Soubeyrand et~al(2009{\natexlab{a}})Soubeyrand, Laine, Hanski, and
  Penttinen}]{soubeyrand_2009_a}
Soubeyrand S, Laine AL, Hanski I, Penttinen A (2009{\natexlab{a}})
  Spatio-temporal structure of host-pathogen interactions in a metapopulation.
  The American Naturalist 174:308--320,
  \urlprefix\url{https://doi.org/doi:10.1086/603624}

\bibitem[{Soubeyrand et~al(2009{\natexlab{b}})Soubeyrand, Neuvonen, and
  Penttinen}]{soubeyrand_2009b}
Soubeyrand S, Neuvonen S, Penttinen A (2009{\natexlab{b}})
  Mechanical-statistical modeling in ecology: from outbreak detections to pest
  dynamics. Bulletin of Mathematical Biology 71:318--338,
  \urlprefix\url{https://doi.org/10.1007/s11538-008-9363-9}

\bibitem[{Soubeyrand et~al(2018)Soubeyrand, de~Jerphanion, Martin, Saussac,
  Manceau, Hendrikx, and Lannou}]{soubeyrand2018inferring}
Soubeyrand S, de~Jerphanion P, Martin O, Saussac M, Manceau C, Hendrikx P,
  Lannou C (2018) {What dynamics underly temporal observations? Application to
  the emergence of {\it Xylella fastidiosa} in {France}: probably not a recent
  story}. New Phytologist \urlprefix\url{https://doi.org/10.1111/nph.15177}

\bibitem[{Spiegelhalter et~al(2002)Spiegelhalter, Best, Carlin, and Van
  Der~Linde}]{spiegelhalter2002bayesian}
Spiegelhalter DJ, Best NG, Carlin BP, Van Der~Linde A (2002) Bayesian measures
  of model complexity and fit. Journal of the Royal Statistical Society: Series
  B (Statistical Methodology) 64:583--639,
  \urlprefix\url{https://doi.org/10.1111/1467-9868.00353}

\bibitem[{Strona et~al(2017)Strona, Carstens, and Beck}]{strona2017network}
Strona G, Carstens CJ, Beck PS (2017) {Network analysis reveals why {\it
  Xylella fastidiosa} will persist in Europe}. {Scientific Reports} 7:71,
  \urlprefix\url{https://doi.org/10.1038/s41598-017-00077-z}

\bibitem[{Turchin(1998)}]{turchin1998quantitative}
Turchin P (1998) Quantitative Analysis of Movement: measuring and modeling
  population redistribution in plants and animals. Sinauer, Sunderland,
  Massachusetts

\bibitem[{Verhulst(1838)}]{verhulst1838notice}
Verhulst PF (1838) Notice sur la loi que la population suit dans son
  accroissement. In: {M}ath{\'e}matique \& sciences humaines, vol 167,
  Quetelet, pp 51--81

\bibitem[{Vermeij(1996)}]{vermeij1996agenda}
Vermeij GJ (1996) An agenda for invasion biology. Biological conservation
  78:3--9

\bibitem[{Weinberger(1978)}]{weinberger1978asymptotic}
Weinberger H (1978) Asymptotic behavior of a model in population genetics. In:
  Nonlinear partial differential equations and applications, Springer, pp
  47--96

\bibitem[{White et~al(2017)White, Bullock, Hooftman, and
  Chapman}]{white_modelling_2017}
White SM, Bullock JM, Hooftman DAP, Chapman DS (2017) Modelling the spread and
  control of xylella fastidiosa in the early stages of invasion in apulia,
  italy. Biological Invasions 19:1825--1837,
  \urlprefix\url{https://doi.org/10.1007/s10530-017-1393-5}

\bibitem[{Wikle(2003{\natexlab{a}})}]{wikle_2003b}
Wikle CK (2003{\natexlab{a}}) Hierarchical {B}ayesian models for predicting the
  spread of ecological processes. Ecology 84:1382--1394

\bibitem[{Wikle(2003{\natexlab{b}})}]{wikle_2003a}
Wikle CK (2003{\natexlab{b}}) Hierarchical models in environmental science.
  International Statistical Review 71:181--199

\end{thebibliography}

\end{document}